\documentclass[preprint,12pt]{elsarticle}
\pdfoutput=1
\journal{Physics of the Dark Universe}

\usepackage{amsmath}
\usepackage{amssymb}
\usepackage{amsfonts}
\usepackage{url}
\usepackage[normalem]{ulem}
\usepackage{xcolor}
\usepackage{graphicx}
\usepackage{hyperref}
\usepackage{bm}

\enlargethispage{\baselineskip}

\hypersetup{
    colorlinks=true,
    urlcolor=blue,
    citecolor=blue,
    linkcolor=blue,
    menucolor=blue,
    anchorcolor=blue,
    filecolor=blue
}

\begin{document}
\begin{frontmatter}

\title{Future targets for light gauge bosons from cosmic strings}

\author[a,b]{Hanyu Cheng}
\ead{chenghanyu@sjtu.edu.cn}
\author[a,b]{Luca Visinelli}
\ead{luca.visinelli@sjtu.edu.cn}

\address[a]{Tsung-Dao Lee Institute (TDLI), No.\ 1 Lisuo Road, 201210 Shanghai, China}
\address[b]{School of Physics and Astronomy, Shanghai Jiao Tong University, \\ Dongchuan Road 800, 200240 Shanghai, China}

\date{\today}

\begin{abstract}
Cosmic strings, theoretical one-dimensional topological defects from the early universe, provide a valuable opportunity for exploring dark matter (DM) production and gravitational wave (GW) emission. Our study investigates the production of gauge bosons and GW emission from cosmic string decay, considering the constraints imposed by cosmological observations. We specifically examine how gauge bosons radiated from strings contribute to DM and dark radiation, with limits set by the observed DM abundance and cosmic microwave background data, respectively. Additionally, we analyze the gravitational wave spectrum of this model across both low and high frequencies. Notably, the spectrum typically follows a $f^{-1/3}$ pattern at high frequencies, beyond the pivot frequency $f_*$. We derive an analytical expression for the frequency $f_*$ and confirm its accuracy through numerical verification. Furthermore, we compare the GW spectrum of this model with the forecasted capabilities of future GW observatories, such as the Einstein Telescope, the Laser Interferometer Space Antenna, the Big Bang Observer, and $\mu$Ares. Finally, we discuss how these considerations impact the model parameters, specifically the gauge boson mass and the energy scale of U(1) symmetry breaking, providing insights into how cosmic strings could enhance our understanding of DM and GW astronomy.
\end{abstract}

\flushbottom
\end{frontmatter}

\section{Introduction}
\label{sec:introduction}

Cosmic strings are hypothetical one-dimensional topological defects believed to have formed during the very early stages of the Universe, potentially as a result of primordial phase transitions~\cite{Kibble:1976sj, Vilenkin:1981bx, Vilenkin:1984ib, Witten:1984eb}. Despite being purely theoretical, cosmic strings are of great interest in cosmology due to their potential influence on the large-scale structure of the Universe and their possible observable effects. For instance, cosmic strings could seed the formation of galaxies and clusters in the early Universe, potentially leaving distinct imprints on the cosmic microwave background and large-scale matter distribution that would provide clues to their existence~\cite{Brandenberger:1990ze, Brandenberger:1993by, Charnock:2016nzm, Fernandez:2020vgi}. Cosmic strings could also explain the baryon asymmetry through leptogenesis~\cite{Bhattacharjee:1982zj} and trigger the formation of primordial black holes~\cite{Hawking:1987bn, Fort:1993zb}.

The search for the fingerprints of the physics related to cosmic strings is of primarily importance for several reasons. In fact, confirming the existence of cosmic strings would teach us about fundamental physics at very high energy otherwise inaccessible in ground-based accelerators. We would also learn about the history of the Universe before Big-Bang nucleosynthesis (BBN), an epoch from which we have no data. Finally, we could learn about the nature of dark matter (DM) and other particles related to the physics beyond the Standard Model (SM).

The evolution of the network of cosmic strings related with the spontaneous breaking of a global symmetry has long been discussed since its early establishment~\cite{Vilenkin:1984ib, Witten:1984eb, Vachaspati:1984gt, Davis:1985pt, Davis:1986xc, Vilenkin:1986ku, Harari:1987ht, Davis:1989nj, Dabholkar:1989ju, Hagmann:1990mj, Allen:1991bk, Battye:1993jv, Davis:1995kk}. A key property of such a {\it global} string network is the development of a scaling regime in which each Hubble volume contains around one long string~\cite{Hindmarsh:1994re, Vilenkin:2000jqa}. The scaling is assured by the continuous production of small string loops through the chopping and reconnection of long strings.

Recent advancements in global string simulations have enabled the study of gravitational wave (GW) emissions from a network of global strings~\cite{Gorghetto:2018myk, Gorghetto:2021fsn, Chang:2021afa}. A significant characteristic of the GW background (GWB) produced by these strings is its emission across a wide range of frequencies as the string network evolves. This distinctive feature sets the GWB from strings apart from other possible sources, such as cosmological phase transitions or inflation, which may also have been active before Big Bang Nucleosynthesis (BBN)~\cite{Caprini:2018mtu, Giovannini:1998bp, Bernal:2019lpc, DEramo:2019tit, Vagnozzi:2022qmc, Caprini:2024ofd}. For this, the GWB predicted from new physics has been vigorously studied to pin down the changes brought in by a modified pre-BBN thermal history, and it is actively sought in experiments~\cite{Cui:2017ufi, Cui:2018rwi, Guedes:2018afo, Gouttenoire:2019kij, Auclair:2019wcv, Ramberg:2019dgi, Ramberg:2020oct, Buchmuller:2021mbb, Ghoshal:2023sfa, Servant:2023tua, Baeza-Ballesteros:2023say, Schmitz:2024hxw, Roshan:2024qnv, Chianese:2024gee}. One important example of a global string network is related to the breaking of the global U(1) Peccei-Quinn symmetry invoked to solve the strong CP problem~\cite{Peccei:1977hh, Peccei:1977ur} and possibly leading to the emission of pseudo-Nambu Goldstone bosons from cosmic strings~\cite{Davis:1986xc, Vilenkin:1986ku, Harari:1987ht, Davis:1989nj, Dabholkar:1989ju, Hagmann:1990mj, Allen:1991bk, Battye:1993jv}.

One alternative possibility involves the breaking of a gauge symmetry, leading to the formation of a cosmic string network composed of {\it gauge} strings. A gauge string solution shares similarities with a global string, although the energy from the gradient term outside the core regime exhibits finite tension concentrated within the string core. While loops with lengths below a critical value predominantly decay through particle emission, larger loops shrink via the release of gravitational waves (GWs), similar to the global case~\cite{Matsunami:2019fss, Baeza-Ballesteros:2024otj}.

The production of gauge bosons from the decay of cosmic strings has gained considerable interest in recent years, as it represents a successful model to produce such particles as dark matter in the early Universe~\cite{Jeannerot:1999yn, Lin:2000qq, Matsuda:2005fb, Cui:2008bd, Kuroyanagi:2012jf, Long:2019lwl, Kitajima:2022lre, Kitajima:2023vre}. Alternative mechanisms may suffer some issues. For example, the production from misalignment requires some degree of fine tuning unless the gauge boson couples non-minimally to gravity~\cite{Arias:2012az, Graham:2015rva, Alonso-Alvarez:2019ixv, Caputo:2021eaa}. Relativistic gauge bosons radiated from strings are also expected to contribute to the dark radiation budget to some degree, in particular this component can be constrained by the considerations over the number of relativistic neutrinos that affect free-streaming at recombination. Other models that have been recently considered involve metastable cosmic strings~\cite{Kibble:1976sj, Vilenkin:1982hm, Antusch:2024ypp}.

In this work, we explore the simultaneous production of gauge bosons and the GWB by a local string network. We explicitly consider the Abelian Higgs model, where the U(1) symmetry predicts the existence of a massive gauge field that could contribute as a cold component of the Universe. We compare analytical approaches from the literature and derive the frequency dependence of the GWB for different regimes. These analytical derivations are useful for numerical tests of the string network. These results have been somewhat discussed in previous literature, see e.g.\ Refs.~\cite{Long:2019lwl, Kitajima:2022lre} where the evolution of an Abelian string network is studied with numerical simulations; here, we have delved deeper into the details of the expressions for the GWB and compiled the derivations of GW abundances across various frequency ranges. For completeness, we also compare our results with the sensitivities of planned GW detectors. We work with units $\hbar=c=1$.

\section{Methods}
\label{sec:methods}

\subsection{The Abelian Higgs Model}

We consider a $U(1)$ symmetry that is broken at the energy scale $v$ by the vacuum expectation value of a complex scalar field $\Phi$ as
\begin{equation}
    \Phi = \left(\rho + \frac{v}{\sqrt{2}}\right)\,e^{i\phi/v}\,,
\end{equation}
so that $\langle\Phi\rangle = v/\sqrt{2}$. Here, $\rho$ and $\phi$ are the radial and angular scalar fields in the decomposition of $\Phi$, respectively. The Abelian Higgs Lagrangian describing the coupling of the complex scalar field with the associated gauge boson $A'_\mu$ is
\begin{equation}
    \label{eq:Lagrangian}
    \mathcal{L} = |D_\mu\Phi|^2 - \frac{1}{4}(F')^{\mu\nu}F'_{\mu\nu} - \lambda_\Phi\left(|\Phi|^2 - \frac{v^2}{2}\right)^2\,,   
\end{equation}
where $F'_{\mu\nu} = \partial_\mu A'_\nu-\partial_\nu A'_\mu$ and $D_\mu = \partial_\mu + i2gA'_\mu$ is the covariant derivative that includes $A'_\mu$ with the coupling $g$. Here, $\lambda_\Phi$ is the dimensionless quartic coupling which, in the numerical results below, is set to $\lambda_\Phi=1$. The masses of the gauge boson and the scalar singlet read
\begin{eqnarray}
    m_{A'} &=& {\sqrt{2}}\,g\,v\,,\label{eq:massAp}\\
    m_\rho &=& \sqrt{2\lambda_\Phi}\,v\,,
\end{eqnarray}
while the field $\phi$ associated with the angular direction remains massless. In the following, we treat $m_{A'}$ and $v$ as independent quantities, so that the coupling $g$ is specified through Eq.~\eqref{eq:massAp}.

The Lagrangian in Eq.~\eqref{eq:Lagrangian} allows for a solution in terms of a cosmic string, a linear concentration of energy of cosmological size which is produced along with a cosmological phase transition in the early Universe~\cite{Nielsen:1973cs}, see also Refs.~\cite{Hindmarsh:1994re, Vilenkin:2000jqa}. The string solution is described below.

\subsection{Gravitational wave emission from cosmic strings}
\label{sec:GWemission}

The equation of motion associated with the theory in Eq.~\eqref{eq:Lagrangian} allows for a solution describing a cosmic string~\cite{Vilenkin:1981bx, Vilenkin:1984ib}. The string configuration is described by a scalar field displaced from its minimum within a narrow region of radius $r \leq 1/m_\rho$ and decaying in the outer region to the horizon scale $r\sim 1/H(t)$. The non-zero spatial gradient in the outer region holds most of the energy density of the field, so that the string tension that arises from the spatial gradient of the scalar field's phase depends logarithmically on the ratio of the ultraviolet (UV) scale $1/m_\rho$ and the infrared (IR) scale $\sim 1/H$ as
\begin{equation}
    \label{eq:mu}
    \mu(t) = \pi\,v^2\,\ln\left(m_\rho / H(t)\right)\,.
\end{equation}
The set of cosmic strings develop into a network at energies below the spontaneous symmetry breaking scale $v$, with strings showing a complex dynamics through mutual intersections and reconnections. Within a Hubble time, cosmic strings intersections form new structures such as loops, kinks and cusps~\cite{Vachaspati:1984gt, Garfinkle:1987yw, Damour:2000wa, Damour:2001bk, Olmez:2010bi} which dissolve by the emission of strong bursts of gravitational waves (GWs).

The energy loss rate of a loop of length $\ell$ is given by
\begin{equation}
    \label{eq:energyrate}
    \frac{{\rm d} E_{\ell}}{{\rm d} t} = -\Gamma_{\rm GW} G \mu^2\,,
\end{equation}
where $\Gamma_{\rm GW}$ is a dimensionless constant that describes the share of energy that a single smooth closed loop releases in GWs and the Newton's constant is $G = 1/m_{\rm Pl}^2$ in terms of Plank's mass $m_{\rm Pl}$. Numerical simulations fix this constant as $\Gamma_{\rm GW} \simeq 50$~\cite{Blanco-Pillado:2013qja}. Oscillating closed string loops are typically the dominant source of GWs from a cosmic string network in the scaling regime. For a string of energy $E_{\ell} = \mu\ell$ and neglecting the time dependence of the string tension in the integration, the length $\ell = \ell(t)$ of a string loop created by the network at time $t_i$ evolves according to
\begin{equation}
    \label{eq:length_GW}
    \ell=\alpha t_i - \Gamma_{\rm GW} G \mu \left(t-t_i\right)\,.
\end{equation}
The first term parametrizes the initial loop size in terms of a fraction $\alpha \simeq 10^{-1}$ of the size of the horizon at formation $t_i$. The second term above describes the shortening of the loop as it emits gravitational radiation.

The frequency of the normal mode $k$ of a string loop is $f_{\text {emit}}=2 k / \ell$ upon emission at time $t$. The contribution of the GW strain to the present spectrum happens at a redshifted frequency
\begin{equation}
    \label{eq:frequency}
    f = \frac{a(t)}{a\left(t_0\right)} \frac{2 k}{\ell},
\end{equation}
that accounts for the correction by the ratio of scale factors $a(t)$ between emission and the current time $t_0$. Inserting Eq.~\eqref{eq:length_GW} into Eq.~\eqref{eq:frequency} allows to express $t_i$ for the GW case only as a function of the detected frequency and the emission time,
\begin{equation}
    \label{eq:tiGW}
    t_{i,{\rm GW}} = \frac{a(t)}{a_0} \frac{2 k}{(\alpha+\Gamma_{\rm GW}G\mu) f}+\frac{\Gamma_{\rm GW}G\mu t}{\alpha+\Gamma_{\rm GW}G\mu}\,,
\end{equation}
where $a_0 = a(t_0)$. The approximation $\alpha+\Gamma_{\rm GW}G\mu \approx \alpha$ holds since we account for sub-Planckian scales $G\mu \propto (v/m_{\rm Pl})^2 \ll 1$. A gravitational wave of frequency $f$ emitted from loops of mode $k$ at time $t$ and observed today only receives the contributions from the loops generated at time $t_i$.

The stochastic GWB depends on the rate of loop production by the cosmic string network. We model this using the velocity-dependent one-scale (VOS) model~\cite{Martins:1995tg, Martins:1996jp, Martins:2000cs}. The energy of the network transferred to the loops is defined as
\begin{equation}
    \rho_{\rm loop}(t) = C_{\rm loop}\,\rho_{\rm str}(t)\,,    
\end{equation}
where $C_{\rm loop}$ parameterizes the fraction of energy transferred from the long string network into large loops and $\rho_{\rm str}$ is the string energy density. Here, we fix $C_{\rm loop} \approx 10\%$ according to recent cosmic string simulations~\cite{Blanco-Pillado:2013qja, Blanco-Pillado:2017oxo}, which implies that the remaining 90\% of the energy is transferred to the kinetic energies of highly-boosted smaller loops. Applying local energy conservation and considering only large loops, this yields a loop formation rate per unit volume at time $t_i$ of
\begin{equation}
    \frac{{\rm d} n_{\text {loop }}}{{\rm d} t_i} = C_{\rm loop}\,\frac{C_{\rm eff}}{\alpha} t_i^{-4}\,.
\end{equation}
The function $C_{\rm eff}$ depends on the redshift scaling of the dominant energy density $\rho$ of the universe and it is obtained within the VOS model as $C_{\rm eff}^{\rm MD} =0.41$ for matter domination (MD) and $C_{\rm eff}^{\rm RD} = 5.5$ for radiation domination (RD)~\cite{Cui:2017ufi}. Here, we interpolate between these results by adopting the expression
\begin{equation}
    C_{\rm eff} = C_{\rm eff}^{\rm RD} + (C_{\rm eff}^{\rm MD}-C_{\rm eff}^{\rm RD})\frac{(t_i / t_{\rm eq})^2}{1+(t_i / t_{\rm eq})^2}\,.
\end{equation}

Considering the GW strain emitted at time $t$ from a string loop that originated at time $t_i < t$, the loop formation rate per unit volume at time $t$ is given by
\begin{equation}
    \label{dn/dti}
    \left(\frac{{\rm d} n_{\text {loop }}}{{\rm d} t_i}\!\right)_t = \frac{{\rm d} n_{\text {loop }}}{{\rm d} t_i} \,\left[\frac{a\left(t_i\right)}{a(t)}\right]^3\,,
\end{equation}
where the ratio of cosmological volumes accounts for the effect of the Universe's expansion between $t_i$ and $t$. The number density spectrum per unit frequency is obtained from Eq.~\eqref{dn/dti} as
\begin{equation}
    \left(\frac{{\rm d} n_{\text {loop}}}{{\rm d} f}\!\right)_t = \left(\frac{{\rm d} n_{\text {loop }}}{{\rm d} t_i}\!\right)_t\,\frac{a(t)}{a\left(t_0\right)} \frac{2 k}{\left(\alpha+\Gamma_{\rm GW}G\mu \right)f^2}\,,
\end{equation}
where the factor $\left({\rm d} t_i/{\rm d} f\right)_t$ is recovered from Eq.~\eqref{eq:tiGW}.

The energy density in GWs per unit frequency and unit time, emitted from loops of mode $k$ at time $t$ and measured today, reads
\begin{equation}
    \label{drho/df}
    \frac{{\rm d}^2 \rho_{\rm GW}^{(k)}}{{\rm d} t\, {\rm d} f} = \left(\frac{{\rm d} n_{\text {loop }}}{{\rm d} f}\right)_t\, P_{\rm GW}^{(k)}\, \left[\frac{a(t)}{a\left(t_0\right)}\right]^4\,,
\end{equation}
where the last factor accounts for the redshift of radiation between emission and today, and the average power (energy per unit time) of GWs emitted by a single loop with mode $k$ is modeled in terms of a function $\Gamma_k$ that accounts for the emission details as~\cite{Vilenkin:1981bx, Turok:1984cn, Vachaspati:1984gt, Burden:1985md, Garfinkle:1987yw}
\begin{equation}
    P_{\rm GW}^{(k)}=\Gamma_k G \mu^2\,.
\end{equation}
We consider the GW emission from cusps, which is described by~\cite{Vachaspati:1984gt, Garfinkle:1987yw, Damour:2000wa, Damour:2001bk, Olmez:2010bi}
\begin{equation}
    \label{eq:normalization}
    \Gamma_k = \frac{\Gamma_{\rm GW}}{\zeta(4/3)\,k^{4 / 3}}\,,
\end{equation}
where $\zeta(4/3) \equiv \sum_k k^{-4/3} \approx 3.60$.

Finally, the present GW density per unit logarithmic frequency, obtained by summing over all harmonic modes $k$, is
\begin{equation}
    \label{eq:GWA}
    \Omega_{\rm GW}(f)=\frac{1}{\rho_c} \sum_k \frac{{\rm d} \rho_{\rm GW}^{(k)}}{{\rm d} \ln f}\,,
\end{equation}
where $\rho_c = 3 H_0^2 / (8 \pi G)$ is the present critical density. The GW energy density per unit logarithmic frequency emitted from the formation of the string network at time $t_F \sim 1/(2v)$ to the present time is given by~\cite{Cui:2007js, Cui:2017ufi, Cui:2018rwi}
\begin{equation}
    \label{eq:drhoGWdlnf}
    \begin{split}
    \frac{{\rm d} \rho_{\rm GW}^{(k)}}{{\rm d} \ln f} =& \frac{2 k}{f} \frac{C_{\rm loop}\, \Gamma_k}{\alpha} \!\int_{t_F}^{t_0} {\rm d}t \frac{C_{\rm eff}}{\alpha+\Gamma_{\rm GW}G\mu}\frac{G \mu^2}{t_{i,{\rm GW}}^4}\\
    &\times \left[\frac{a(t)}{a(t_0)}\right]^5\left[\frac{a(t_{i,{\rm GW}})}{a(t)}\right]^3\Theta\left(t-t_{i,{\rm GW}}\right)\,,
    \end{split}
\end{equation}
where $t_{i,{\rm GW}}$ is the formation time obtained from Eq.~\eqref{eq:tiGW} and $\Theta$ is the Heaviside theta function. When the string network emits only GWs, mimicking a global string network, the GW abundance in Eq.~\eqref{eq:GWA} at high frequencies predicts $\Omega_{\rm GW}(f) \propto \sqrt{\mu}$~\cite{Blanco-Pillado:2013qja}.

\subsection{The role of gauge bosons emission}
\label{sec:GB}

The string network forming within the Abelian Higgs model might simultaneously decay by emitting gauge bosons. This additional decay channel has significant implications for the evolution of the network. The energy loss rate of a loop of length $\ell$ is given by
\begin{equation}
    \label{eq:energyrateGB}
    \frac{{\rm d} E_{\ell}}{{\rm d} t} = -\Gamma_{\rm GW} G \mu^2-\Gamma_{\rm vec} v^2\Theta\left(1-m_{A'}\, \ell\right)\,,
\end{equation}
where the dimensionless constant $\Gamma_{\rm vec} \sim 65$ parameterizes the strength of the release in gauge bosons, with this numerical value determined by numerical simulations~\cite{Blanco-Pillado:2013qja}. The step function in Eq.~\eqref{eq:energyrate} ensures that the loop cannot emit gauge bosons if the inverse of the initial loop size is smaller than the gauge boson mass. The length scale $1/m_{A'}$ introduces a new cutoff scale, which could potentially be smaller than the horizon scale, thereby modifying the string tension in Eq.~\eqref{eq:mu} as~\cite{Gorghetto:2018myk}
\begin{equation}
    \label{eq:muGB}
    \mu = \pi\,v^2\,\log\,,
\end{equation}
where the last term depends on the cutoff energies at both the UV and IR scales through the constant $M \equiv \max \left[H, H_{A'} \right]$ as
\begin{equation}
    \label{eq:log}
    \log = \ln\left(m_\rho / M\right)\,.
\end{equation}
Here, the term $H_{A'} \equiv \alpha m_{A'}/2$ describes the Hubble rate at which the effects of the gauge boson mass become relevant. The corresponding critical time is
\begin{equation}
    t_{A'} \equiv (\alpha m_{A'})^{-1}\,,
\end{equation}
where we have included an additional factor of $\alpha$ with respect to previous literature for consistency with Eq.~\eqref{eq:energyrateGB}.

The inclusion of the emission into gauge bosons modifies the rate at which string loops shrink in the purely gravitational case. In deriving Eq.~\eqref{eq:length_GW} it has been assumed that the time dependence of the string tension does not affect the result. This agrees with the form of the time dependent term in Eq.~\eqref{eq:mu} which is a logarithmic correction that receives the largest contribution around times $t \lesssim t_{A'}$, with the contributions from early times affecting the numerical value by less than $\mathcal{O}(1)$. In the code, the expression in Eq.~\eqref{eq:length_GW} is used to derive the value for the emission time $t_i$ as a function of $t$. The expression assures that when $t_i<t_{A'}$ is satisfied, the rate $\Gamma_{\rm eff}(t)$ accounts for the emission in both GWs and gauge bosons, while for $t_i > t_{A'}$ only the GW emission is present. This occurs because the size of a cosmic string loop formed at times $t_i > t_{A'}$ is larger than $H^{-1}(t_{A'})$, forcing $\Theta\left(1-m_{A'} \ell\right)=0$ and the dynamics driven solely by GW emission at later times.

We calculate the present GW density per unit logarithmic frequency by considering the contribution from both large loops $m_{A'} \ell_i>1$, which only emits GWs, and small loops $m_{A'} \ell_i<1$, which decay by emitting both GWs and gauge bosons. For this, we modify Eq.~\eqref{eq:length_GW} by including this additional decay channel, so that the length of a string loop emitted at $t_i$ results in
\begin{equation}
    \label{eq:length_GB}
    \ell=\alpha t_i- \Gamma_{\rm eff}(t) \left(t-t_i\right)\,.
\end{equation}
Here, the term $\Gamma_{\rm eff}(t)$ describes the shortening of the loop as it emits both gravitational radiation and gauge bosons, with a different expression depending on the moment at which the gauge boson is released. In fact, defining the critical length scale $\ell_{A'} = 1/m_{A'}$ using Eq.~\eqref{eq:energyrateGB}, for early times $t \leq t_{A'}$ it is $\log = \ln(m_\rho/H(t))$ in Eq.~\eqref{eq:log}, while for late times $t>t_{A'}$ it is $\log = \ln(2m_\rho\,t_{A'})$, so that
\begin{equation}
    \label{eq:Gammaeff}
    \Gamma_{\rm eff}(t) \equiv \Gamma_{\rm GW} G \mu + \Gamma_{\rm vec}/\left(\pi \log\right)\,.
\end{equation}
This expression accounts for the fact that, when loops emit both GW and gauge boson, the loop length evolves with the time according to Eq.~\eqref{eq:length_GB} with the function $\Gamma_{\rm eff}(t)$ in place of the constant $\Gamma_{\rm GW}$. With the definition in Eq.~\eqref{eq:Gammaeff}, the formation time of the loop in Eq.~\eqref{eq:tiGW} modifies as
\begin{equation}
    \label{eq:tiGB}
    t_i = \frac{a(t)}{a_0} \frac{2 k}{(\alpha+\Gamma_{\rm eff}(t)) f}+\frac{\Gamma_{\rm eff}(t)t}{\alpha+\Gamma_{\rm eff}(t)}\,.
\end{equation}
The GW energy density per unit log frequency modifies from the expression in Eq.~\eqref{eq:drhoGWdlnf} to
\begin{equation}
    \label{eq:drhoGWdlnfGB}
    \begin{split}
    \frac{{\rm d} \rho_{\rm GW}^{(k)}}{{\rm d} \ln f} &= \frac{2 k}{f} \frac{C_{\rm loop}\, \Gamma_k}{\alpha} \!\int_{t_F}^{t_0} {\rm d}t \frac{C_{\rm eff}}{\alpha+\Gamma_{\rm eff}(t)}\frac{G \mu^2}{t_i^4}\\
    &\times \left[\frac{a(t)}{a(t_0)}\right]^5\left[\frac{a(t_i)}{a(t)}\right]^3\Theta\left(t-t_i\right)\,.
    \end{split}
\end{equation}
This expression reduces to the spectrum for the emission of GWs only when setting $\Gamma_{\rm eff}(t) \approx \Gamma_{\rm GW}G\mu$ and for a constant string tension in Eq.~\eqref{eq:muGB}. The loop length evolves over time according to Eq.~\eqref{eq:length_GB}, which accounts for the emission of both gravitational waves (GWs) and gauge bosons by the loop. We can then calculate the GW density per unit logarithmic frequency observed today for small loops ($m_{A'} \ell_i < 1$) that emit both GWs and gauge bosons similarly. For large loops ($m_{A'} \ell_i > 1$), only GWs are emitted. For a given GW frequency $f$ observed today from mode $k$, the emission time $t$ is related to the loop formation time through Eq.~\eqref{eq:tiGB}.

Solving Eq.~\eqref{eq:drhoGWdlnfGB} allows for the reconstruction of the GW spectrum as a function of frequency. In particular, when compared with numerical results (see, e.g., Ref.~\cite{Kitajima:2023vre}), our method reproduces the same pivot frequencies and spectral slopes of the GW spectrum, although it generally predicts peak amplitudes that are smaller by about an order of magnitude.

\subsection{Relic gauge boson abundance}
\label{sec:DM}

The evolution of the string energy density is governed by the release of gauge boson emission at early times, according to the expression~\cite{Harari:1987ht}
\begin{equation}
    \dot \rho_{\rm str} + 2H\rho_{\rm str} = -\mathcal{P}\,,
\end{equation}
where a dot is a differentiation with respect to time $t$ and the power spectrum density $\mathcal{P}$ is the total amount of energy radiated from the string network per unit time and volume,
\begin{equation}
    \mathcal{P} \approx \frac{\xi(t) \mu(t)}{t^3}\,.
\end{equation}
Here, $\xi$ is the average string number per Hubble volume, which appears to show a logarithmic violation of the scaling law as $\xi = \xi_0 + \xi_1\ln(m_\rho/H)$ in some numerical simulations~\cite{Kawasaki:2018bzv, Vaquero:2018tib, Gorghetto:2018myk, Buschmann:2019icd, Klaer:2019fxc, Gorghetto:2021fsn, Buschmann:2021sdq}, with other studies claiming the conventional scaling law $\xi \approx 1$~\cite{Yamaguchi:1998gx, Yamaguchi:1999yp, Yamaguchi:1999dy, Hagmann:2000ja, Hiramatsu:2010yu, Hiramatsu:2012gg, Kawasaki:2014sqa, Lopez-Eiguren:2017dmc, Hindmarsh:2019csc, Hindmarsh:2021vih}. Here, we adopt the former approach and consider the average number of long strings per Hubble volume showing a logarithmic dependence on the gauge boson mass as $\xi = \xi_1 \ln(m_\rho/m_{A'})$, where we fix $\xi_1 = 0.15$.

The quanta of the gauge field emitted by the string network add up to the cosmological energy budget, contributing to the total dark matter abundance for relatively large masses $m_{A'}$. The number density of gauge boson evolves according to
\begin{equation}
    \label{num_den_evo}
    \dot n_{A'} + 3 H n_{A'} = \mathcal S_{A'}\,,
\end{equation}
where a dot is a derivation with respect to cosmic time $t$ and the number of gauge boson produced per unit time and volume at time $t$ in terms of the power density $\mathcal{P}$ and the energy of the particle $E_k=\sqrt{k^2+m_{A'}^2}$ as
\begin{equation}
    \mathcal{S}_{A'} = \int_0^{\infty} \mathrm{d} k \frac{1}{E_k} \frac{{\rm d} \mathcal{P}}{{\rm d} k}\,.
\end{equation}
Here, the integration runs over the magnitude of the momentum of the radiated particles $k$. In the assumption that only the modes with energy $E_k \sim H(t)$ contribute, the source density reads $\mathcal{S}_A \approx \mathcal{P} / H(t)$, so that the solution of Eq.~\eqref{num_den_evo} leads to the present abundance of non-relativistic gauge bosons as~\cite{Kitajima:2022lre}
\begin{equation}
    \label{eq:DMabundance}
    \Omega_{A'} h^2 \simeq 0.091\left(\frac{\xi}{12}\right)\left(\frac{m_{A'}}{10^{-13} \,\mathrm{eV}}\right)^{1 / 2}\left(\frac{v}{10^{14} \,\mathrm{GeV}}\right)^2\,.
\end{equation}
Note, that alternative derivations in the literature find the same power-law dependence, despite differing by an overall numerical prefactor~\cite{Gorghetto:2018myk, Long:2019lwl}. 

Along with the cold component, the network emits relativistic gauge bosons that add up to the dark radiation budget in the cosmic microwave background (CMB). The contribution at times $t < t_{A'}$ leads to~\cite{Gorghetto:2021fsn}
\begin{equation}
    \label{rho_A_until_t_*}
    \rho_{A'}(t) = \frac{4}{3} \xi \mu H^2 \log\,,
\end{equation}
while at later times, radiation redshifts as $a^{-4}$. In the standard cosmological scenario, Eq.~\eqref{rho_A_until_t_*} holds until matter-radiation equality since $H \propto a^{-2}$ during radiation domination, so that the energy density in gauge bosons at CMB is
\begin{equation}
    \rho_{A'}(t_{\rm CMB}) = \frac{64 \pi^2 G \xi_1 v^2 \log ^3}{9}\left(\frac{g_*(t_{\rm eq})}{g_\gamma}\right) \rho_\gamma(t_{\text {CMB}})\,.
\end{equation}
Here, $g_\gamma = 2$ accounts for the photon polarizations and $g_*(t_{\rm eq}) = 3.36$. Using the contribution from neutrinos as
\begin{equation}
    \rho_\nu(t_{\text {CMB}}) = \frac{7}{8}\left(\frac{4}{11}\right)^{4 / 3} \rho_\gamma(t_{\text {CMB}})\,,
\end{equation}
the excess in neutrino degrees of freedom at CMB is
\begin{equation}
    \label{eq:Neff}
    \begin{split}
    \Delta N_{\rm eff}&=\frac{\rho_{A'}(t_{\rm CMB})}{\rho_\nu(t_{\rm CMB})}\\
    &= \frac{512 \pi^2}{63}\left(\frac{11}{4}\right)^{4 / 3}\left(\frac{g_*(t_{\rm eq})}{g_\gamma}\right)\,Gv^2 \xi_1\log ^3\,.
    \end{split}
\end{equation}
In principle, Eq.~\eqref{eq:Neff} should be modified to account for the effects of a non-zero mass at CMB~\cite{Caloni:2022uya}. We do not consider such results here because the bound on $m_{A'}$ resulting from Eq.~\eqref{eq:DMabundance} imposes that the gauge boson be lighter than the temperatures in play at CMB.

\subsection{Cosmological setup}

To derive our results below, we fix the cosmological parameters according to the best fit by the {\it Planck} collaboration for the TT,TE,EE+lowE+lensing datasets~\cite{Planck:2018vyg}. More in detail, the standard cosmological model parameters are set by taking the Hubble constant as $H_0 = 100\,h{\rm\,km\,s^{-1}\,Mpc^{-1}}$ with $h = 0.674$. The redshift at matter-radiation equality is approximately $z_{\rm eq} \approx 3402$. The matter density parameter is fixed as $\Omega_m = 0.3153$, leading to a radiation density parameter of $\Omega_r = \Omega_m/(1+z_{\rm eq}) = 9.3\times10^{-5}$. The dark energy density parameter is then $\Omega_\Lambda \approx 0.685$.  The scale factor as a function of cosmic time can be obtained by integrating the Friedmann equation
\begin{equation}
    \frac{1}{a}\frac{{\rm d}a}{{\rm d}t} = H_0\,\left(\Omega_m a^{-3} + \Omega_\Lambda+\Omega_r a^{-4}\right)^{1/2}\,.
\end{equation}
This integration also provides the age of the Universe, which, given the chosen cosmological parameters, is approximated as $t_0 = 0.951\,H_0^{-1}$.

For the analytical derivation below, we model the scale factor as
\begin{equation}
    a(t) \simeq
    \begin{cases}
        a_{\rm eq} (t/t_{\rm eq})^{1/2}\,,&\quad\hbox{for $t\leq t_{\rm eq}$ or RD}\,,\\
        a_0 (t/t_0)^{2/3}\,,&\quad\hbox{for $t> t_{\rm eq}$ or MD}\,,
    \end{cases}
\end{equation}
where the scale factor at matter-radiation equality is $a_{\rm eq} = (1+z_{\rm eq})^{-1}$ and the time at matter-radiation equality is fixed as
\begin{equation}
    t_{\rm eq}^{-1} \approx \frac{3}{2}H_0\sqrt{2\Omega_m}a_{\rm eq}^{-3/2}\,.
\end{equation}
These expressions are used for the estimates provided below.

\section{Results}
\label{sec:results}

\subsection{GW emission}
\label{sec:GWemissionresults}

We first investigate the behavior of the GW abundance as described by Eq.~\eqref{eq:GWA}, focusing on the GW emissions from cosmic strings generated during either the matter domination or radiation domination epochs separately. The GW spectrum given in Eq.~\eqref{eq:drhoGWdlnf} can be schematically represented as
\begin{equation}
    \frac{{\rm d} \rho_{\rm GW}^{(k)}}{{\rm d} \ln f} = \frac{{\rm d} \rho_{\rm GW}^{(k)}}{{\rm d} \ln f}\bigg|_{\rm RD,RD} \!\!+ \frac{{\rm d} \rho_{\rm GW}^{(k)}}{{\rm d} \ln f}\bigg|_{\rm RD,MD} \!\!+ \frac{{\rm d} \rho_{\rm GW}^{(k)}}{{\rm d} \ln f}\bigg|_{\rm MD,MD}\,,
\end{equation}
where the first term is the contribution during RD of strings formed during RD, the second term is the contribution during MD from strings formed during RD, and the third term is the contribution at MD from strings formed during MD.

We first consider the case of GW emission during RD. For a constant string tension, Eq.~\eqref{eq:drhoGWdlnf} can be integrated exactly in the limit of high frequencies and for radiation domination,
\begin{equation}
    (1+z_{\rm eq})(\Gamma_{\rm GW} G\mu\,\sqrt{t\,t_{\rm eq}})\,f \gg 1\,,
\end{equation}
to provide the expression
\begin{equation}
    \label{eq:drhoGWdlnf_GW}
    \frac{{\rm d} \rho_{\rm GW}^{(k)}}{{\rm d} \ln f}\bigg|_{\rm RD,RD} \approx \frac{4 a_{\rm eq}^4\,C_{\rm eff} C_{\rm loop}}{3Gt_{\rm eq}^2}  \frac{\Gamma_k}{\Gamma_{\rm GW}}\sqrt{\frac{\alpha\,G\mu}{\Gamma_{\rm GW}}}\,.
\end{equation}
The expression obtained in this limit is independent of time and frequency, and it depends on the integer $k$ only through $\Gamma_k$, as shown in Eq.~\eqref{eq:normalization}. After performing the summation in Eq.~\eqref{eq:GWA}, we find
\begin{equation}
    \label{eq:GWA1}
    \begin{split}
    \Omega_{\rm GW}(f) &\approx 16\pi\, C_{\rm eff} C_{\rm loop}\,\Omega_m a_{\rm eq}\,\sqrt{\frac{\alpha\,G\mu}{\Gamma_{\rm GW}}}\\
    &\approx 8\times 10^{-4}\sqrt{\frac{G\mu}{\Gamma_{\rm GW}}}\,.
    \end{split}
\end{equation}
The expression above predicts that, in the radiation-dominated cosmology, the GW abundance at high frequencies is constant and depends on the string tension as $\Omega_{\rm GW}(f) \propto \sqrt{\mu}$~\cite{Blanco-Pillado:2013qja}. At the opposite end, for low frequencies we find
\begin{equation}
    \label{eq:GWA2}
    \begin{split}
    \Omega_{\rm GW}(f) &\approx 2\,C_{\rm eff} C_{\rm loop} \sqrt{\alpha} \Gamma_{\rm GW} (G \mu)^2\left(\Omega_m\,a_{\rm eq}^7\right)^{\frac{1}{4}}\left(\!\frac{f}{H_0}\!\right)^{\frac{3}{2}}\\
    &\approx 2\times 10^{-7}\,\Gamma_{\rm GW}\,(G \mu)^2\,\left(\frac{f}{H_0}\right)^{\frac{3}{2}}\,.
    \end{split}
\end{equation}

\begin{figure}[ht!]
    \centering
    \includegraphics[width=\linewidth]{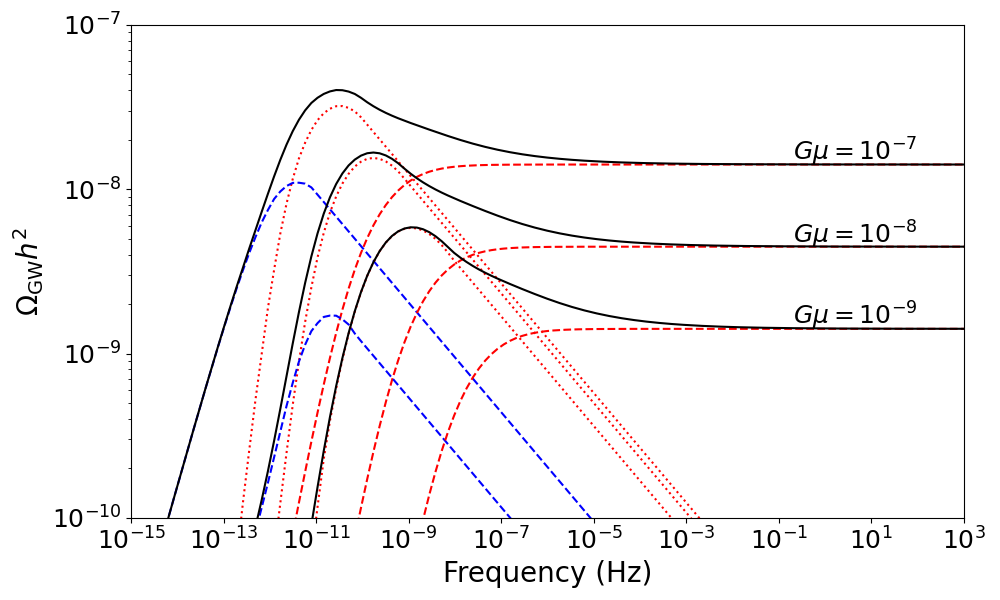}
    \caption{The abundance of primordial GWs from cosmic strings for different values of the string tension $G\mu$. Results are for the GWs produced from loops produced and emitting during RD (red dashed lines), produced during RD and emitting during MD (red dotted lines), produced and emitting during MD (blue dotted lines), and the overall results (black solid lines).}
    \label{fig:spectrum_only_GW1}
\end{figure}

We now consider the case where a loop formed during RD, requiring $t_i < t_{\rm eq}$, and we restrict the analysis to the GW emission during MD or $t > t_{\rm eq}$. In this scenario, Eq.~\eqref{eq:drhoGWdlnf} becomes
\begin{equation}
    \label{eq:drhoGWdlnfRDMD}
    \frac{{\rm d} \rho_{\rm GW}^{(k)}}{{\rm d} \ln f}\bigg|_{\rm RD,MD}\!\!\!\!= \frac{2 k}{f} \frac{C_{\rm loop}C_{\rm eff}\Gamma_k}{\alpha(\alpha+\Gamma_{\rm GW}G\mu)} \!\int_{t_{\rm eq}}^{t_m}\!\! {\rm d}t \frac{G \mu^2}{t_i^{5/2}}\frac{a_{\rm eq}^3}{t_{\rm eq}^{3/2}}\left(\frac{t}{t_0}\right)^{\frac{4}{3}}\!\!,
\end{equation}
where the formation time from Eq.~\eqref{eq:tiGW} is
\begin{equation}
    \label{eq:tiMD}
    t_i = \left(\frac{t}{t_0}\right)^{\frac{2}{3}}\frac{2 k}{(\alpha+\Gamma_{\rm GW}G\mu) f}+\frac{\Gamma_{\rm GW}G\mu t}{\alpha+\Gamma_{\rm GW}G\mu}\,.
\end{equation}
In Eq.~\eqref{eq:drhoGWdlnfRDMD}, the largest emission time $t_m$ of GWs coincides with $t_0$, provided the frequencies are high enough to satisfy the condition $t_i < t_{\rm eq}$, ensuring that the loop forms during RD. In this context, the integration of Eq.~\eqref{eq:drhoGWdlnfRDMD} at low frequencies yields the expression
\begin{equation}
    \label{eq:RDMD_smallf}
    \begin{split}        
    \Omega_{\rm GW} &\approx 2\,C_{\rm eff} C_{\rm loop} \sqrt{\alpha} \Gamma_{\rm GW} (G \mu)^2\left(\Omega_m\,a_{\rm eq}^9\right)^{\frac{1}{4}}\left(\frac{t_0}{t_{\rm eq}}\right)\left(\!\frac{f}{H_0}\!\right)^{\frac{3}{2}}\\
    &\approx 10^{-3}\,\Gamma_{\rm GW} (G \mu)^2\,\left(\frac{f}{H_0}\!\right)^{\frac{3}{2}}\,.
    \end{split}
\end{equation}
This predicts the behavior $\Omega_{\rm GW} \propto f^{3/2}$, as shown by the red dotted lines in Fig.~\ref{fig:spectrum_only_GW1}. For even lower frequencies, the constraint $t_i < t_{\rm eq}$ affects the upper limit of integration in Eq.~\eqref{eq:drhoGWdlnfRDMD}, which then effectively becomes
\begin{equation}
    t_m = t_0\,\min\left[1, \left(\frac{\alpha\,t_{\rm eq}\,f}{2k}\right)^{3/2}\right]\,.
\end{equation}
We find that $t_m < t_0$ for $f < 2k/(\alpha\,t_{\rm eq}) \approx 10^{-11}\,k\,\text{Hz}$, which represents a critical frequency. Below this frequency, the next-to-leading term in the expansion series of Eq.~\eqref{eq:drhoGWdlnfRDMD} becomes more significant, resulting in a change in the behavior of the spectrum to $\Omega_{\rm GW} \propto f^{5/2}$. This condition remains valid down to the time $t_m = t_{\rm eq}$, which corresponds to the minimum possible frequency of the emission,
\begin{equation}
    f > \frac{2k}{\alpha\,t_{\rm eq}}\,\left(\frac{t_{\rm eq}}{t_0}\right)^{2/3} \approx 3\times10^{-15}\,k{\rm\,Hz}\,.
\end{equation}

For high frequencies, Eq.~\eqref{eq:drhoGWdlnfRDMD} predicts the behavior $\Omega_{\rm GW} \propto f^{-1}$ when considering the first few modes. However, the slope changes to $\Omega_{\rm GW} \propto f^{-1/3}$ when the summation over a large number of modes is included. This change in behavior can be estimated by expanding Eq.~\eqref{eq:drhoGWdlnfRDMD} to second order in $f^{-1}$ and comparing the leading contribution with the next-to-leading term, yielding the pivot frequency
\begin{equation}
    f_* \approx \frac{5}{3\,\Gamma_{\rm GW}\,G\mu\,t_0}\left(\frac{t_0}{t_s}\right)^{\frac{1}{3}} \frac{1 - \left(\frac{t_s}{t_0}\right)^{\frac{1}{2}}}{1 - \left(\frac{t_s}{t_0}\right)^{\frac{1}{6}}}\,,
\end{equation}
where $t_s$ labels the earliest emission time of GWs during MD. For the case where only GW emission is considered, $t_s = t_{\rm eq}$, the expression above simplifies to
\begin{equation}
    \label{eq:fstarRDMD}
    f_* \approx 130\,\left(\Gamma_{\rm GW}\,G\mu\,t_0\right)^{-1}\,,
\end{equation}
which leads to a good fit with the numerical resolution of Eq.~\eqref{eq:drhoGWdlnfRDMD}. Indeed, the numerical integration for $f > f_*$ yields $\Omega_{\rm GW} \propto f^{-1/3}$.

We finally deal with the case of emission at MD for a loop forming during MD, for which Eq.~\eqref{eq:drhoGWdlnf} reads
\begin{equation}
    \label{eq:drhoGWdlnfMDMD}
    \frac{{\rm d} \rho_{\rm GW}^{(k)}}{{\rm d} \ln f}\bigg|_{\rm MD,MD} = \frac{2 k}{f} \frac{C_{\rm loop}C_{\rm eff}\Gamma_k}{\alpha^2} \!\int_{t_{\rm low}}^{t_0} {\rm d}t \frac{G \mu^2}{t_i^2t_0^2}\left(\frac{t}{t_0}\right)^{\frac{4}{3}},
\end{equation}
for the same expression as in Eq.~\eqref{eq:tiMD} for the formation time, and $t_{\rm low}$ is defined so that $t_{\rm i}(t_{\rm low}) \approx t_{\rm eq}$. In the limit of low frequencies, this expression gives
\begin{equation}
    \Omega_{\rm GW} \approx 2\,C_{\rm loop} C_{\rm eff}\, \Gamma_{\rm GW}\, (G\mu)^2\,t_0\,f\,,
\end{equation}
predicting the behavior $\Omega_{\rm GW} \propto f$ at low frequencies, see the blue dashed lines in Fig.~\ref{fig:spectrum_only_GW1}. Likewise, the behavior at high frequencies modifies from $\Omega_{\rm GW} \propto f^{-1}$ to $\Omega_{\rm GW} \propto f^{-1/3}$ when summing over a large number of modes. This occurs at frequencies above the pivot frequency
\begin{equation}
    f_* \approx 16\,\left(\Gamma_{\rm GW}\,G\mu\,t_0\right)^{-1}\,.
\end{equation}

As a test, we reproduce Figure~13 from Ref.~\cite{Blanco-Pillado:2013qja} for string tensions $G\mu = 10^{-7}$, $G\mu = 10^{-8}$, and $G\mu = 10^{-9}$. The results are shown in Fig.~\ref{fig:spectrum_only_GW1} for the GWs produced by loops formed and emitting during RD (red dashed lines), formed during RD and emitting during MD (red dotted lines), formed and emitting during MD (blue dotted lines), and the overall results (black solid lines). We find a $\mathcal{O}(1)$ discrepancy between the analytical expression in Eqs.~\eqref{eq:GWA1} and~\eqref{eq:GWA2} and the numerical result, which arises from the definition of $t_{\rm eq}$.

\subsection{Including gauge boson emission}

We extend the discussion of GW emission from a cosmic string network in Sec.~\ref{sec:GWemission} to include the impact of gauge boson dynamics. Since we are considering gauge boson masses significantly above $1/t_{\rm eq} \sim 10^{-27}\,$eV, their additional emission primarily affects the spectrum at high frequencies, which corresponds to loops formed during RD.

The term describing the loop production and GW emission during RD is
\begin{equation}
    \label{eq:drhoGWdlnfRDRDhighf}
    \begin{split}
    &\frac{{\rm d} \rho_{\rm GW}^{(k)}}{{\rm d} \ln f}\bigg|_{\rm RD,RD}\\
    \approx&{} \frac{2 k}{f}\frac{C_{\rm eff}C_{\rm loop}}{t_{\rm eq}^2}\frac{\Gamma_k}{\alpha^2} \frac{a_{\rm eq}^5}{t_{\rm eq}^{1/2}} \int_{t_s}^{t_{\rm eq}}\!\!\frac{{\rm d} t\,t\,G \mu^2}{\left(a_{\rm eq}\left(\frac{t}{t_{\rm eq}}\right)^{\frac{1}{2}} \frac{2 k}{f \alpha}+\frac{\Gamma_{\rm GW} G \mu}{\alpha} t\right)^{\frac{5}{2}}}\\
    \approx&{} \frac{4\alpha^{1/2}}{3}\Gamma_k\frac{C_{\rm eff}C_{\rm loop}}{t_{\rm eq}^2}a_{\rm eq}^{4}\,G \mu^2 \\
    &\times\!\bigg[\!\!\left(\frac{a_{\rm eq}}{t_{\rm eq}} \frac{2 k}{f} + \Gamma_{\rm GW} G \mu\right)^{\!-\frac{3}{2}} \!\!\!\!\!-\! \left(\frac{a_{\rm eq}}{\sqrt{t_s \,t_{\rm eq}}} \frac{2 k}{f} + \Gamma_{\rm GW} G \mu\right)^{\!-\frac{3}{2}}\!\!\bigg],
    \end{split}
\end{equation}
where $t_s$ is defined so that $t_i(t_s) \approx t_{A'}$, or
\begin{equation}
    t_s \approx (\Gamma_{\rm GW}\,G \mu\,m_{A'})^{-1}\,.
\end{equation}
where the approximation holds at high frequencies.

For $k=1$, the expression in Eq.~\eqref{eq:drhoGWdlnfRDRDhighf} predicts the behaviour $\Omega_{\rm GW} \propto f^{-1}$ at high frequencies. However, the inclusion of higher modes modifies the frequency scaling of the GW abundance to $\Omega_{\rm GW} \propto f^{-1/3}$~\cite{Blasi:2020wpy, Gouttenoire:2019kij, Cui:2019kkd}. We have verified that the numerical results obtained through Eq.~\eqref{eq:GWA} exhibit this behavior at high frequencies in the limit of a large number of modes. This same behaviour can also be derived analytically as follows: to recover the correct behavior at high frequencies, we approximate the summation over $k$ with an integral and take the limit at large $k$. The resulting expression is independent of the mode and reads
\begin{equation}
    \label{eq:drhoGWdlnfRDRDhighf1}
    \Omega_{\rm GW}^{\rm RD,RD} \approx 15\,\frac{C_{\rm eff}C_{\rm loop}}{H_0^2}\frac{a_{\rm eq}^4}{t_{\rm eq}^2}\sqrt{\frac{\alpha\,G\mu}{\Gamma_{\rm GW}}}\left(\frac{a_{\rm eq}m_{A'}}{f}\right)^{\frac{1}{3}}\xi\,,
\end{equation}
where we set $\xi = (\Gamma_{\rm GW}\,G \mu\,t_{\rm eq}\,m_{A'})^{-1/6}$. The expression above shows the expected behavior $\Omega_{\rm GW}^{\rm RD,RD} \propto f^{-1/3}$ at high frequencies. Eqs.~\eqref{eq:GWA1} and~\eqref{eq:drhoGWdlnfRDRDhighf1} match at the pivot frequency 
\begin{equation}
    \label{eq:fstar}
    f_* \approx \frac{2.4}{t_0} \left(\frac{\left(1+z_{\rm eq}\right)\,t_{\rm eq}}{\alpha \Gamma_{\rm GW} G \mu t_{A'}}\right)^{1/2}\,.
\end{equation}
This expression matches the findings in Ref.~\cite{Blasi:2020wpy} up to a numerical prefactor.

For loops formed at RD and emitting during MD, Eq.~\eqref{eq:drhoGWdlnfRDMD} is modified when accounting for the solution of the equation $t_i(t_s) \approx t_{A'}$, with $t_i$ given by Eq.~\eqref{eq:tiMD}. This leads to
\begin{equation}
    \label{eq:fstarRDMD1}
    f_* \approx \frac{5 m_{A'}}{3}\,\beta\,\frac{1 - \beta^{\frac{1}{2}}}{\beta^{\frac{1}{3}}-\beta^{\frac{1}{2}}}\,,
\end{equation}
with $\beta = (\Gamma_{\rm GW}\,G \mu\,t_0\,m_{A'})^{-1}$. The actual pivot frequency is the minimum value between Eq.~\eqref{eq:fstarRDMD} and Eq.~\eqref{eq:fstarRDMD1}.

\begin{figure}[htbp]
    \centering
    \includegraphics[width=\linewidth]{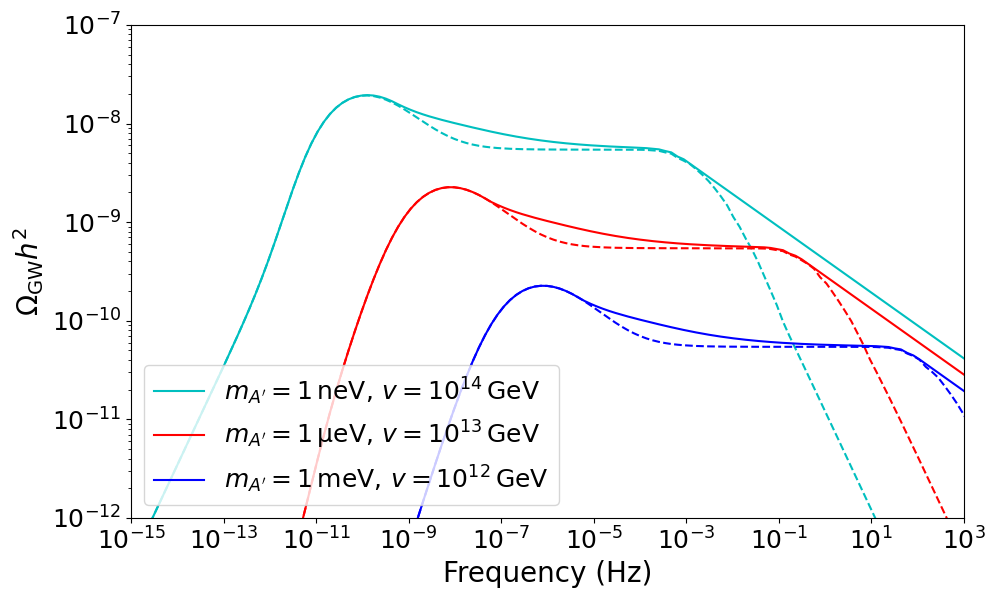}
    \caption{The GW spectrum obtained from Eq.~\eqref{eq:GWA} considering the first $k_{\rm tot} = 15$ modes (dashed lines) and using the methods described in Eqs.~\eqref{eq:drhoGWdlnfRDRDhighf} and~\eqref{eq:drhoGWdlnfRDRDhighf1} (solid lines), for the parameters combinations $(m_{A'}, v)$ = $(1{\rm\,neV}, 10^{14}{\rm\,GeV})$ (cyan), $(1{\rm\,\mu eV}, 10^{13}{\rm\,GeV})$ (red), and $(1{\rm\,meV}, 10^{12}{\rm\,GeV})$ (blue).}
    \label{fig:comparison}
\end{figure}

In Fig.~\ref{fig:comparison}, we compare the spectra obtained from Eq.~\eqref{eq:GWA} considering the first $k_{\rm tot} = 15$ modes (dashed lines) with those obtained using the methods described above (solid lines), for the parameters combinations $(m_{A'}, v)$ = $(1{\rm\,neV}, 10^{14}{\rm\,GeV})$ (cyan), $(1{\rm\,\mu eV}, 10^{13}{\rm\,GeV})$ (red), and $(1{\rm\,meV}, 10^{12}{\rm\,GeV})$ (blue). The approximations used for the solid lines are consistent with the evaluation of Eq.~\eqref{eq:GWA} for a large number of modes, $k_{\rm tot} \gg 1$.

\section{Discussion}
\label{sec:discussion}

\begin{figure}[htbp]
    \centering
    \includegraphics[width=\linewidth]{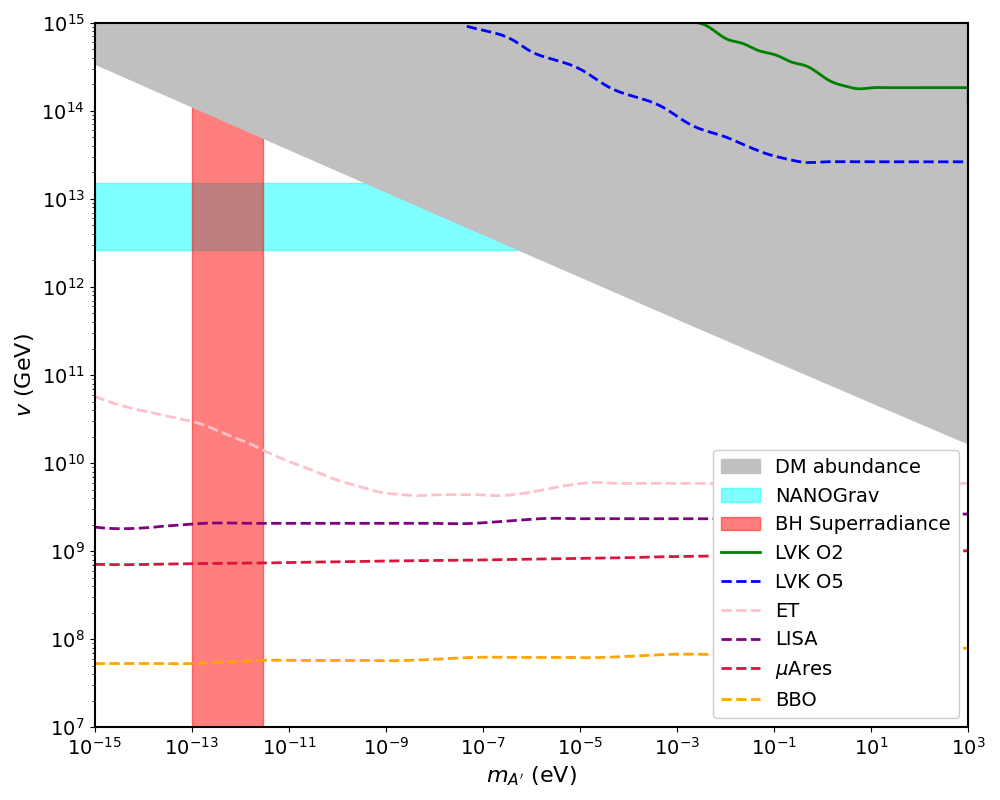}
    \caption{Constraints on the model parameters $m_{A'}$ and $v$, assuming all other parameters in the emission as described in the text. The gray area is excluded due to the over-production of gauge bosons as the DM. The red shaded vertical band is excluded by considerations from BH superradiance. The cyan shaded horizontal band is hinted by the measurements of the GWB by the NANOGrav collaboration. The LVK O2 has already excluded the top right corner of the plot (green solid line). The forecast reach for other interferometers is shown, namely LVK O5 (blue), ET (pink), LISA (purple), $\mu$Ares (red), and BBO (orange).}
    \label{fig:interferometerbounds}
\end{figure}

We assess the model against the forecast reach of various GW measurements and the bounds obtained from cosmological considerations. Fig.~\ref{fig:interferometerbounds} shows the results in terms of the model parameters: the gauge boson mass $m_{A'}$ (horizontal axis) and the energy scale $v$ (vertical axis). The parameter space is constrained in the upper right portion by the requirement that the contribution from cold gauge bosons produced via the misalignment mechanism does not overclose the Universe~\cite{Long:2019lwl, Kitajima:2023vre}. This constraint is derived by enforcing $\Omega_{\rm DM}h^2 \lesssim 0.12$ in Eq.~\eqref{eq:DMabundance} (gray shaded area). Note that the result shown in Fig.~\ref{fig:interferometerbounds} below includes the estimate for DM abundance reported in Ref.~\cite{Kitajima:2022lre}, which differs from the analysis in Ref.~\cite{Long:2019lwl} by a factor $\sim\mathcal{O}(1)$. This discrepancy likely arises from unresolved differences between the simulations, particularly concerning the abundance of transverse vector bosons present when the network dissipates. We will not delve further into this discussion here, as our goal is to highlight the potential target for future experiments. Another upper bound, not shown here, comes from the requirement that the contribution of relativistic gauge bosons to the effective number of neutrinos does not exceed the limit allowed by CMB measurements, see Eq.~\eqref{eq:Neff}. For the mass range considered here, this bound is not relevant, though it would become significant for lighter gauge bosons.

We perform a scan over the parameter space to determine whether, for a given set of values $(m_{A'}, v)$, the corresponding GW strain falls within the forecasted reach of a given GW interferometer. We include the reach of LIGO-Virgo-Kagra (LVK) during run 2 (O2) and run 5 (O5)~\cite{LIGOScientific:2007fwp, KAGRA:2013rdx, LIGOScientific:2014pky, LIGOScientific:2016fpe}, represented by the green solid and blue dashed lines.

We also consider the forecasts from other GW experiments. The Einstein Telescope (ET) is a proposed ground-based GW observatory expected to operate in the Hz-kHz frequency bands, with a peak strain sensitivity of $h \sim \mathcal{O}(10^{-25}){\rm\,Hz^{-1/2}}$~\cite{Punturo:2010zz, Sathyaprakash:2012jk, Maggiore:2019uih}. Since ET will operate within the frequency bands covered by LVK interferometers but with improved sensitivity, the forecasted region for ET shows a similar shape at lower values of $v$. The region that can be explored with ground-based interferometry weakens at low masses due to the GW spectrum cutting off at light masses, corresponding to the frequency cutoff in Eq.~\eqref{eq:fstar}.

At lower frequencies, we account for the forecast reach of the upcoming Laser Interferometer Space Antenna (LISA) satellite~\cite{Amaro-Seoane:2012vvq, Barausse:2020rsu, LISACosmologyWorkingGroup:2022jok, Blanco-Pillado:2024aca}, which has been recently funded by the European Space Agency (ESA) and is planned for launch in the 2030s. This is represented by the area bounded by the dashed purple line in Fig.~\ref{fig:interferometerbounds}. A proposed successor to LISA by ESA is the Big Bang Observer (BBO), which aims to probe the mHz to kHz frequency range, peaking at $f_{\rm GW}\sim 1\,$Hz with a strain sensitivity of $|h|\sim 10^{-24}{\rm\,Hz^{-1/2}}$~\cite{Yagi:2011wg}. BBO will extend the search for cosmic strings from a phase transition well below the current limits, reaching down to the symmetry-breaking scale $v \lesssim 10^8\,$GeV. 

At even lower frequencies, the proposed $\mu$Ares experiment aims to cover the GW bands in the $\mu$Hz to mHz frequency range~\cite{Sesana:2019vho}. An overview of the expected sensitivities for GW interferometers can be found in Refs.~\cite{Schmitz:2020syl, schmitz_2020_3689582}. The results of the scan for all the considered interferometer forecasts are shown in Fig.~\ref{fig:interferometerbounds}, with specific curves for LVK O2 (green), LVK O5 (blue), ET (pink), LISA (purple), $\mu$Ares (red), and BBO (orange). Besides LVK which has already concluded its run O2 and it is represented by a solid line, all expected forecast reaches are shown with dashed lines. Since BBO, LISA and $\mu$Ares operate below the mHz range, these interferometers are unable to probe the pivot frequency which scales with the string energy density as $f_* \propto (G\mu)^{-1}$. Consequently, they cannot determine the nature of the string network (global versus gauge). The results for LISA align with those obtained for $\alpha=0.1$ in Ref.~\cite{Auclair:2019wcv}.

The scaling at extremely low frequencies, corresponding to the nHz range, could potentially contribute to the detection of the GWB through pulsar timing analysis. Earlier work constrains the string tension as $G\mu \lesssim 8\times 10^{-10}$ using data from the European Pulsar Timing Array (EPTA)~\cite{EPTA:2015qep}, $G\mu \lesssim 5.3\times 10^{-11}$ using data from North American Nanohertz Observatory for Gravitational Waves (NANOGrav)~\cite{NANOGRAV:2018hou}, or $G\mu \lesssim 5.1\times 10^{-10}$ using data from Parkes Pulsar Timing Array (PPTA)~\cite{Chen:2022azo}. The recent detection of the GW spectrum at such low frequencies suggests a string tension in the range $-11 \lesssim \log_{10}G\mu \lesssim -9.5$, with the exact values depending on the specific distribution of loops in the string network~\cite{EPTA:2023xxk, NANOGrav:2023hvm}. This is illustrated by the cyan shading in Fig.~\ref{fig:interferometerbounds}. Also shown is the constraint from BH superradiance~\cite{Arvanitaki:2009fg, Arvanitaki:2010sy} derived from the observation of highly spinning black holes (red dashed area), which restricts the mass of the gauge boson as $10^{-13}{\rm\,eV} \lesssim m_{A'} \lesssim 2\times 10^{-11}{\rm\,eV}$~\cite{Baryakhtar:2017ngi, Cardoso:2018tly, Mehta:2021pwf}. We restrict ourselves to the results obtained for masses $m_{A'} \gtrsim 10^{-15}$\,eV for a few reasons. Firstly, lighter bosons in the mass range $10^{-19}{\rm\,eV} \lesssim m_{A'} \lesssim 10^{-16}{\rm\,eV}$ are disfavored by the observation of spinning supermassive black holes at various mass ranges above $10^6\,M_\odot$~\cite{Brito:2014wla, Mehta:2021pwf}. Heavier black holes with masses in the range $\gtrsim 10^9\,M_\odot$ could even probe the mass scale below $10^{-21}$\,eV~\cite{Davoudiasl:2019nlo, Chen:2022nbb}, although such massive BHs lack precise measurements of their spins. Besides, various cosmological measurements disfavor a large portion of the DM being in the form of a light boson, such as the Lyman-$\alpha$ absorption lines caused by neutral hydrogen in the intergalactic medium and constraining to $m_{A'}\gtrsim 10^{-21}$\,eV~\cite{Hui:2016ltb, Kobayashi:2017jcf, Armengaud:2017nkf, Winch:2024mrt}.

We do not expect the gauge boson mass to be ultralight in models that connect the dark photon to the Standard Model. The mass of the light boson is generally related to the vacuum energy scale $v$ as $m_{A’} \sim g v$, where $g$ is the gauge boson coupling. In models based on gauged symmetries, such as baryon number conservation, the gauge boson is expected to couple with neutrinos, implying that $v$ is linked to beyond the Standard Model physics at a high energy scale. The gauge boson coupling is also loosely constrained by the weak gravity conjecture~\cite{Arkani-Hamed:2006emk}, which requires $g \gtrsim m_\nu / m_{\rm Pl} \sim 10^{-30}$ for a neutrino mass $m_\nu \sim 10^{-2}$\,eV. For example, setting $v \gtrsim 10^{8}$\,GeV, which is the lowest energy considered in Fig.~\ref{fig:interferometerbounds}, leads to $m_{A’} \gtrsim 10^{-13}$\,eV. Although model building allows for theories with a gauge boson mass spanning various orders of magnitude~\cite{Agrawal:2018vin, Dror:2018pdh, Bastero-Gil:2018uel, Co:2018lka, Visinelli:2024wyw}, we do not consider ultralight bosons here.

As anticipated in the discussion around Fig.~\ref{fig:interferometerbounds}, we have incorporated additional bounds from complementary sources. Constraints on the number of effective degrees of freedom $N_{\rm eff}$ have been used to place an upper limit on the SSB scale $v$ due to the production of relativistic gauge bosons. These measurements also constrain the abundance of the stochastic GW signal at various frequencies, with stricter bounds applied at higher frequencies. This constraint impacts the amount of GWs produced by cosmic strings of cosmological origin, as these GWs would be present during BBN and at CMB decoupling. At BBN, this results in~\cite{Caprini:2018mtu}
\begin{equation}
    \label{eq:rhoBBN}
    \frac{\rho_{\rm GW}}{\rho_\gamma} < \frac{7}{8}\left(\frac{4}{11}\right)^{4/3}\,\Delta N_{\rm eff}\,,
\end{equation}
where $\rho_{\rm GW}$ refers to the portion of the GWB that is present before BBN occurs. Following Eq.~\eqref{eq:drhoGWdlnfGB}, the energy density in GWs is~\cite{Servant:2023tua}
\begin{equation}
    \rho_{\rm GW} = \sum_k\,\int {\rm d}\ln f\,\frac{{\rm d} \rho_{\rm GW}^{(k)}}{{\rm d} \ln f}\,,
\end{equation}
where the integration runs over the cutoff frequencies of the background. We modify Eq.~\eqref{eq:drhoGWdlnf} to account for the loops formed to BBN as
\begin{equation}
    \label{eq:drhoGWdlnfBBN}
    \begin{split}
    \frac{{\rm d} \rho_{\rm GW}^{(k)}}{{\rm d} \ln f} =& \frac{2 k}{f} \frac{C_{\rm loop}\, \Gamma_k}{\alpha} \!\int_{t_F}^{t_{\rm BBN}} {\rm d}t \frac{C_{\rm eff}}{\alpha+\Gamma_{\rm GW}G\mu}\frac{G \mu^2}{t_i^4}\\
    &\times \left[\frac{a(t)}{a(t_{\rm BBN})}\right]^5\left[\frac{a(t_i)}{a(t)}\right]^3\Theta\left(t-t_i\right)\,,
    \end{split}
\end{equation}
where $t_i$ is computed assuming that frequencies redshift till $t_{\rm BBN}$. The result leads to $\rho_{\rm GW} \propto f^{3/2}$ up to frequencies $f \lesssim 0.1{\rm\,Hz}/G\mu$, which is much larger than the maximum frequency emitted at BBN $f_{\rm max}$. In fact, setting the temperature scale $T_{\rm BBN} = 1\,$MeV gives $f_{\rm max} \sim 1/t_{\rm BBN} \sim 2\,$Hz, so that the integration in Eq.~\eqref{eq:rhoBBN} gives
\begin{equation}
    \label{eq:GWBBN}
    \rho_{\rm GW} \approx 0.16\,C_{\rm eff} C_{\rm loop} \Gamma_{\rm GW}\alpha^{1/2}\left(ft_{\rm BBN}\right)^{3/2}\frac{G\mu^2}{t_{\rm BBN}^2}\,.
\end{equation}
We adopt the recent results for $\Delta N_{\rm eff}$ from the {\it Planck} satellite~\cite{Planck:2018vyg},
\begin{equation}
    \label{eq:Neffbound}
    \Delta N_{\rm eff} \lesssim 0.376\,,\hbox{at 95\% confidence level}\,,
\end{equation}
and the parameters considered here leads to the bound
\begin{equation}
    v \lesssim 2\times 10^{15}{\rm\,GeV}\,,
\end{equation}
which is of the same order of magnitude as what is obtained from considering the emission of relativistic gauge bosons.

\begin{figure}[ht!]
    \centering
    \includegraphics[width=\linewidth]{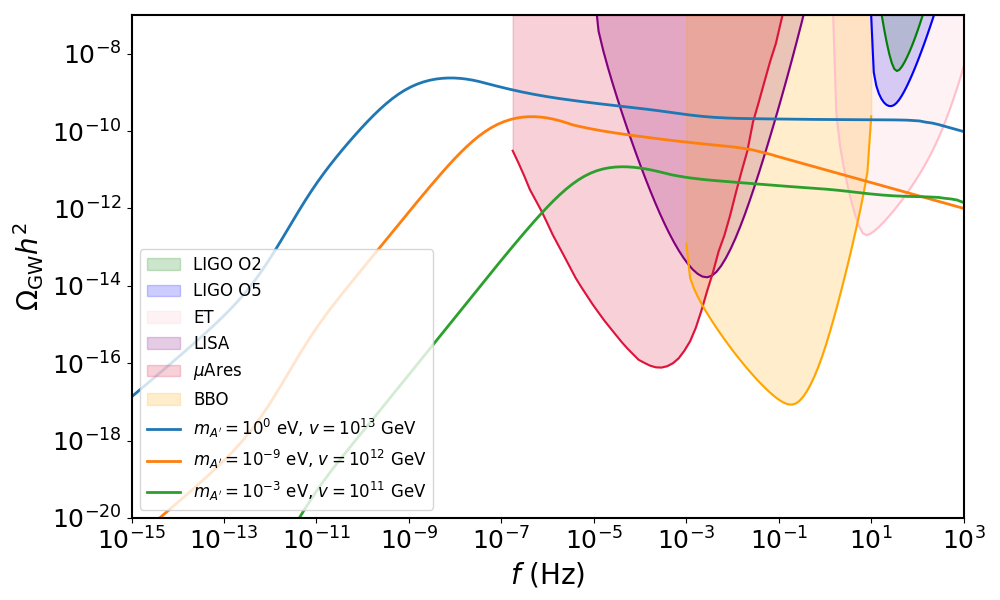}
    \caption{GW spectrum from a cosmic string network as a function of frequencies. We choose the model parameters ($m_{A'}, v$) = (1\,eV, 10$^{13}$\,GeV) (blue line), ($m_{A'}, v$) = (10$^{-9}$\,eV, 10$^{12}$\,GeV) (orange line), ($m_{A'}, v$) = (10$^{-3}$\,eV, 10$^{11}$\,GeV) (green line). The Run 2 of LIGO-Virgo has excluded the top right corner of the plot (green solid line). Also shown is the forecast reach for LVK O5 (blue), ET (pink), LISA (purple), $\mu$Ares (red), and BBO (orange).}
    \label{fig:spectrum}
\end{figure}

To exemplify the procedure, in Fig.~\ref{fig:spectrum} we compare the reach for the experimental setups considered with the predictions from the GWB from the string network for some combinations of the parameters ($m_{A'}, v$), namely ($m_{A'}, v$) = (1\,eV, 10$^{13}$\,GeV) (blue), ($m_{A'}, v$) = (10$^{-9}$\,eV, 10$^{12}$\,GeV) (orange), ($m_{A'}, v$) = (10$^{-3}$\,eV, 10$^{11}$\,GeV) (green). For each choice, the resulting spectrum is shown as a function of the GW frequency at detection.

\section{Conclusions}

We have examined the gravitational wave (GW) background predicted in Abelian Higgs models, assessing the region of parameter space that will be within the reach of third-generation gravitational-wave observatories. Our work advances previous literature by providing a detailed analytical treatment of the expressions used and discussing the impact of massive gauge boson emission on the GW spectrum. We assess the model in the context of the forecasted capabilities of future interferometers, including LISA, BBO, ET, $\mu$Ares, and LIGO O5, and compare these with recent results from NANOGrav. The gauge boson emission is constrained by considerations for dark radiation and dark matter abundances. Fundamental physical parameters such as the energy scale of a primordial phase transition in the early Universe and the mass of a new light gauge boson can be inferred from reconstructing the GW spectrum at different frequencies.

For all of these proposed or planned interferometers, we perform a scan over the parameters $(m_{A'}, v)$ to decide which region of the parameter space could be covered with future missions. The results of this analysis are summarized in Fig.~\ref{fig:interferometerbounds}, following the comparisons exemplified in Fig.~\ref{fig:spectrum}. Future work on this topic will revolve around the understanding of the GW spectrum in scenarios where the Abelian Higgs model and its extensions are relevant. This includes, but is not limited to, exploring the implications of the $B\textrm{--}L$ gauge symmetry, which plays a crucial role in many grand unified theories~\cite{Wilczek:1979hc} and physics beyond the Standard Model~\cite{Escudero:2018fwn, Lin:2022xbu}.

\section*{Acknowledgements}
We thank Michael Zantedeschi for interesting conversations and suggestions on the topic. The authors acknowledge support by the National Natural Science Foundation of China (NSFC) through the grant No.\ 12350610240 ``Astrophysical Axion Laboratories''. L.V.\ thanks for the hospitality received by the Istituto Nazionale di Fisica Nucleare (INFN) section of Ferrara (Italy), the Trento Institute For Fundamental Physics And Applications (TIFPA), the INFN Frascati National Laboratories (Italy), and the Texas Center for Cosmology and Astroparticle Physics, Weinberg Institute for Theoretical Physics at the University of Texas at Austin (USA) for hosting and for the support throughout the completion of this work. This publication is based upon work from the COST Actions ``COSMIC WISPers'' (CA21106) and ``Addressing observational tensions in cosmology with systematics and fundamental physics (CosmoVerse)'' (CA21136), both supported by COST (European Cooperation in Science and Technology).

\bibliographystyle{elsarticle-num}
\bibliography{references.bib}

\end{document}